\RequirePackage{snapshot}

\documentclass[sigplan,10pt]{acmart}
\acmConference[AGERE'18]{ACM SIGPLAN International Workshop on Programming Based on Actors,
Agents, and Decentralized Control}{November 05--07, 2018}{Boston, MA, USA}
\acmYear{2018}
\acmISBN{} 
\acmDOI{} 
\startPage{1}

\setcopyright{none}

\usepackage{booktabs}   
\usepackage{subcaption} 

\usepackage{amssymb,amsmath}
\hypersetup{
  pdfkeywords={session types; concurrency; forwarding; linking}
  pdftitle={Implementing Linking in Multiparty Sessions},
  pdfauthor={Hanwen Wu, Hongwei Xi},
}

\usepackage{url}

\usepackage{mathtools}

\usepackage{todonotes}

\usepackage{pgfmath}
\usepackage{soul}        

\usepackage{tikz}
\usepackage{tikz-cd}
\usetikzlibrary{arrows,shapes,petri,decorations.pathmorphing}
\tikzcdset{arrows={line width=1.414*rule_thickness}}

\usepackage{listings}
\usepackage[capitalise,noabbrev,nameinlink]{cleveref} 
\crefname{sec}{Section}{Sections}
\crefname{prop}{Proposition}{Propositions}
\crefname{lemma}{Lemma}{Lemmas}
\crefname{def}{Definition}{Definitions}
\crefname{thm}{Theorem}{Theorems}
\crefname{app}{Appendix}{Appendices}
\crefname{eg}{Example}{Examples}
\crefname{lst}{Listing}{Listings}
\crefname{remark}{Remark}{Remarks}

\usepackage{ifxetex,ifluatex}
\usepackage{fixltx2e} 
\ifnum 0\ifxetex 1\fi\ifluatex 1\fi=0 
  \usepackage[T1]{fontenc}
  \usepackage[utf8]{inputenc}
\else 
  \ifxetex
    \usepackage{mathspec}
  \else
    \usepackage{fontspec}
  \fi
  \defaultfontfeatures{Ligatures=TeX,Scale=MatchLowercase}
\fi
\IfFileExists{upquote.sty}{\usepackage{upquote}}{}

\begin{document}
%

\newcommand{\rulename}[1]{\text{{\upshape\bf #1}}} 
\newcommand{\ptitle}[1]{{\bf #1}\kern.5em}
\def\ATS{\mathcal{A}\mathcal{T}\kern-.1em\mathcal{S}}
\def\zmq{\O{}MQ}
\def\defeq{\Coloneqq}  

\def\Lzero{\lambda_0}
\def\Lpi{\lambda_0^{\pi}}
\def\Ldep{\lambda_{\forall,\exists}}
\def\Ldeppi{\lambda_{\forall,\exists}^{\pi}}

\newcommand{\sta}[1]{{\color{blue!60!black}{#1}}}

\def\sortt(#1){\sta{{\color{black}{\text{\rm\it #1}}}}} 
\def\sortm(#1){\sta{{\color{black}{#1}}}}               

\def\sorttc(#1){\text{\rm\it #1}}

\def\stset{\sortt(set)}
\def\stint{\sortt(int)}
\def\stbool{\sortt(bool)}
\def\sttype{\sortt(type)}
\def\stvtype{\sortt(vtype)}
\def\ststype{\sortt(stype)}
\def\strole{\sortt(role)}
\def\stsigma{\sortm(\sigma)}
\def\stnat{\sortt(nat)}

\def\stsetc{\text{\rm\it set}}
\def\stintc{\text{\rm\it int}}
\def\stboolc{\text{\rm\it bool}}
\def\sttypec{\text{\rm\it type}}
\def\stvtypec{\text{\rm\it vtype}}
\def\ststypec{\text{\rm\it stype}}
\def\strolec{\text{\rm\it role}}
\def\stnatc{\text{\rm\it nat}}

\newcommand{\stermt}[1]{\sta{\text{\rm\it #1}}\kern0.1em} 
\newcommand{\stermm}[1]{\sta{#1}}               
\newcommand{\stermtc}[1]{\text{\rm\it #1}}      
\newcommand{\stermmc}[1]{#1}      				

\def\subtype{\mathbin{\stermm{\leq_{ty}}}}
\def\sassert{\mathbin{\stermm{\wedge}}}
\def\sguard{\mathbin{\stermm{\supset}}}

\def\sr{\stermm{r}}
\def\snegr{\stermm{\neg r}}

\def\strue{\stermm{\top}}
\def\sfalse{\stermm{\bot}}

\def\rs{\stermtc{rs}}   
\def\rsfull{\rs_{\fullset}}
\def\scx{\stermt{scx}}
\def\scc{\stermt{scc}}
\def\scf{\stermt{scf}}


\def\typet(#1){\stermt{#1}}  
\def\typem(#1){\stermm{#1}}
\def\typetc(#1){\stermtc{#1}}  
\def\typemc(#1){\stermmc{#1}}

\newcommand{\chan}[2]{\typem({\typet(chan)({#1},\stypem({#2}))})}
\newcommand{\chanc}[2]{\typemc({\typetc(chan)({#1},\stypemc({#2}))})}

\def\sunit{\typet(unit)} 
\def\sunitc{\typetc(unit)}
\def\sint{\typet(int)}
\def\sintc{\typetc(int)}
\def\sintn(#1){\sta{\sint(\stermm{#1})}}
\def\sstring{\typet(string)}
\def\sstringc{\typetc(string)}
\def\sbool{\typet(bool)}
\def\sboolb(#1){\sta{\sbool(\stermm{#1})}}
\def\stau{\typem(\tau)}
\def\shattau{\typem(\hat\tau)}

\def\stypet(#1){{\color{cyan}{\texttt{#1}}}}  
\def\stypem(#1){{\color{cyan}{#1}}} 		  
\def\stypemc(#1){{#1}} 		  

\def\stype{\stypet} 
\def\spi{\stypem(\pi)}

\def\snd{\stypet(snd)}
\def\rcv{\stypet(rcv)}
\def\msg{\stypet(msg)}
\def\seqs{\mathbin{\stypem(\dblcolon)}}
\def\quan{\stypet(quan)}
\def\fix{\stypet(fix)}
\def\nil{\stypet(end)}
\def\branch{\stypet(branch)}
\def\ite{\stypet(ite)}
\def\init{\stypet(init)}

\newcommand{\dyn}[1]{{\color{red!50!black}{#1}}}

\def\dkw(#1){\text{\rm\bf #1}}

\def\dlam{\dkw(lam)~}    
\def\dfix{\dkw(fix)~}    
\def\dapp{\dkw(app)}     
\def\dfst{\dkw(fst)}     
\def\dsnd{\dkw(snd)}     
\def\din{~\dkw(in)~}     
\def\dif{\dkw(if)~}      
\def\dthen{~\dkw(then)~} 
\def\delse{~\dkw(else)~} 

\newcommand{\dtermt}[1]{\dyn{\text{\rm #1}}} 
\newcommand{\dtermm}[1]{\dyn{#1}}            

\def\dtrue{\dtermt{true}}
\def\dfalse{\dtermt{false}}
\def\dpair(#1,#2){\dtermm{\langle{#1},{#2}\rangle}}           
\def\dunit{\dtermm{\langle\rangle}}                           
\newcommand{\dite}[3]{\dtermm{\dif{#1}\dthen{#2}\delse{#3}}}  
\newcommand{\dlet}[3]{\dtermm{\dkw(let)~{#1}={#2}\din{#3}}}   

\def\dguardi{\dtermm{\supset^+}}  
\def\dguarde{\dtermm{\supset^-}}
\def\dassert{\dtermm{{\wedge}}}
\def\dforalli{\dtermm{\forall^+}}
\def\dforalle{\dtermm{\forall^-}}
\def\dexists{\dtermm{{\exists}}}

\def\dcx{\dtermt{dcx}}
\def\dcc{\dtermt{dcc}}
\def\dcf{\dtermt{dcf}}
\def\dcr{\dtermt{dcr}}
\def\dxf{\dtermm{\text{\rm\it xf}}}
\def\dtid{\dtermt{tid}}
\def\dpool{\dtermm{\Pi}}


\def\dapi(#1){\textrm{\color{red}{#1}}}

\def\dfork{\dapi(fork)}
\def\dsend{\dapi(send)}
\def\drecv{\dapi(recv)}
\def\dbsend{\dapi(bsend)}
\def\dbrecv{\dapi(brecv)}
\def\dskip{\dapi(skip)}
\def\dclose{\dapi(close)}
\def\dwait{\dapi(wait)}
\def\dcut{\dapi(link)}
\def\dappend{\dapi(append)}
\def\dunify{\dapi(unify)}
\def\dexify{\dapi(exify)}
\def\delim{\dapi(elim)}
\def\dsplit{\dapi(split)}
\def\drecurse{\dapi(recurse)}
\def\doffer{\dapi(offer)}
\def\dchoose{\dapi(choose)}
\def\daccept{\dapi(accept)}
\def\drequest{\dapi(request)}


\def\res{{\color{black}{\mathcal{R}}}}      
\def\sig{{\color{black}{\mathcal{S}}}}      
\def\ectx{{\color{black}{E}}}               
\def\cctx{{\color{black}{C}}}               

\def\map(#1,#2){[{#1}\mapsto{#2}]}         
\newcommand{\subst}[2]{[\sfrac{#2}{#1}]}   


\def\reduce{\mathrel{\longrightarrow}}        
\def\reduceall{\mathrel{\longrightarrow^*}}
\def\vreduce{\stackrel{v}{\longrightarrow}}   
\def\creduce{\stackrel{P}{\reduce}}           
\def\creduceall{\mathrel{\overset{P}{\reduce}\!\!^*}}
\def\dfreduce{\mathrel{\leadsto}}             
\def\dfreducevia(#1){\stackrel{\dtermm{{#1}}}{\dfreduce}} 
\def\dfnormal{\not\dfreduce}
\newcommand{\redex}[1]{\underline{#1}}


\def\channels{\textsf{\textup{channels}}}
\def\endpoints{\textsf{\textup{endpoints}}}
\def\roles{\textsf{\textup{roles}}}

\def\fv{\textit{fv}\kern0.1em}        
\def\fn{\textit{fn}\kern0.1em}        
\def\dom{\textit{dom}\kern0.1em}      

\def\D{\mathscr{D}}         
\def\height{\textit{ht}}    

\def\M{\mathcal{M}}        


\def\ploop{\textsf{\textup{loop}}}
\def\pblocked{\textsf{\textup{blocked}}}
\def\pregular{\textsf{\textup{regular}}}
\def\pdfreducible{\textsf{\textup{df-reducible}}}
\def\pnormal{\textsf{\textup{df-normal}}}
\def\pdeadlocked{\textsf{\textup{deadlocked}}}
\def\pconsistent{\textsf{\textup{consistent}}}
\def\pmatch{\textsf{\textup{match}}}
\def\prelaxed{\textsf{\textup{relaxed}}}


\def\tensor{\varotimes}
\def\parr{\bindnasrepma}
\def\with{\binampersand}

\def\f{{\mathscr{F}}}
\def\uf{{\mathscr{U}}}
\def\conju{\uf}
\def\forallu{\uf^{\lambda}}
\def\mconju{\uf^\times}
\def\aconju{\uf^+}
\def\expou{\uf^\ast}
\def\impliesu{\uf^f}

\def\fullset{\overline{\varnothing}}
\newcommand{\inv}[1]{{#1}^{\raisebox{.2ex}{$\scriptscriptstyle-\!1$}}}

\newcommand{\iform}[2]{[{#1}]_{{#2}}}
\newcommand{\iformc}[2]{\cp{\iform{#1}{#2}}}


\def\cli{\stermm{C}}
\def\srv{\stermm{S}}

\title[]{Implementing Linking in Multiparty Sessions}         
\subtitle{(Extended Abstract)}                     



\author{Hanwen Wu}
\affiliation{
  \department{Dept. of Computer Science}              
  \institution{Boston University}            
  \streetaddress{111 Cummington Mall}
  \city{Boston}
  \state{MA}
  \postcode{2215}
  \country{USA}                    
}
\email{hwwu@bu.edu}          
\author{Hongwei Xi}
\affiliation{
  \department{Dept. of Computer Science}              
  \institution{Boston University}            
  \streetaddress{111 Cummington Mall}
  \city{Boston}
  \state{MA}
  \postcode{2215}
  \country{USA}                    
}
\email{hwxi@bu.edu}          


\begin{abstract}
The fast growth of service-oriented programming (SOP) is evident in this
day and age of the Internet, and handling communication is of paramount
importance in SOP. Session types are a formalism that is proposed to
specify interactions between communicating processes, where the word
``session'' loosely refers to a (possibly infinite) sequence of such
interactions. In essence, a session type system is a kind of type system
designed to enforce (through type-checking) that the involved processes
communicate according to a chosen protocol specified as a session type.
It is well-known that linear logic plays a pivotal role in the study of
session types. For instance, various inference rules in linear logic can
be interpreted as ways for constructing channels (used by communicating
processes to send/receive messages.) A particularly interesting case is
the cut-rule in linear logic, which can be interpreted as a way for
connecting the ends of two matching channels to form a single new
channel. This form of channel construction is often referred to as
linking or (bi-directional) forwarding.

\par

We have generalized classical linear logic into classical linear
multirole logic (LMRL), where the former can be seen as a special case
of the latter involving only two roles. In LMRL, there is a cut-rule
involving multiple sequents (instead of exactly two), which we call
multiparty cut (mp-cut). We have also formulated a novel multiparty
session type system directly based on LMRL. When implementing it, we
need to find a way of connecting multiple channels that corresponds to
mp-cut.

\par

In this paper, we describe an implementation of linking for multiparty
sessions in the setting of shared memory. We also describe two novel
concepts, two-way linking with \emph{residual} and \emph{three-way}
linking, which can only be formulated in the setting of multiparty
sessions. Notably, linking for binary sessions can be thought of as a
specially optimized version of what is implemented for multiparty
sessions.
\end{abstract}



\maketitle

\hypertarget{introduction}{%
\section{Introduction}\label{introduction}}

Original session types \cite{Honda:1993eh,Honda:1998fm,Takeuchi:1994bv}
are binary in the sense that they are formulated for specifying
communication protocols between exactly \emph{two parties}, which are
connected via a \emph{channel} with two \emph{endpoints}, usually held
by some \emph{threads} implementing the parties. As an example, let us
assume that two programs \(P\) and \(Q\) are connected with a
bi-directional channel. We may think of \(P\) as a client who sends two
integers to the server \(Q\) and then receives from \(Q\) either true or
false depending on whether or not the first sent integer is less than
the second one. The communication protocol can be described using a
session type of the following form:
\[\stypem({\msg(P,\stintc)\seqs\msg(P,\stintc)\seqs\msg(Q,\stboolc)\seqs\nil(P)})\]
which means that an integer is to be sent from \(P\), another integer is
to be sent from \(P\), a boolean is to be sent from \(Q\), and finally
the channel is to be closed by \(P\). The session type system will
ensure \(P\) and \(Q\) implement the protocol \emph{dually} from their
respective local perspective. The session between \(P\) and \(Q\) is
bounded in the sense that it contains only a bounded number of sends and
receives. By introducing recursively defined session types, unbounded
sessions containing indefinite numbers of sends and receives can be
readily specified.

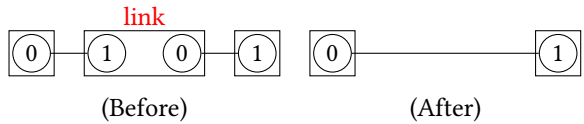
\begin{figure}
\begin{tikzpicture}
\tikzstyle{ep}=[circle,draw,minimum size=0.5cm,inner sep=0]
\tikzstyle{proc}=[rectangle,draw,minimum size=0.6cm,inner sep=0]

\begin{scope}
\node[ep] (a0) at (-1.5, 0.0) {0};
\node[ep] (a1) at (-0.5, 0.0) {1};
\node[ep] (b0) at (0.5, 0.0) {0};
\node[ep] (b1) at (1.5, 0.0) {1};
\draw (a0) -- (a1);
\draw (b0) -- (b1);
\node[proc] (p0) at (-1.5,0.0) {};
\node[proc] (q1) at (1.5,0.0) {};
\draw (-0.8,0.3) rectangle (0.8,-0.3);
\node [above] at (0,0.25) {$\dcut$};
\node at (0,-0.75) {(Before)};
\end{scope}

\begin{scope}[xshift=4cm]
\node[ep] (a0) at (-1.5, 0.0) {0};
\node[ep] (b1) at (1.5, 0.0) {1};
\node[proc] (p0) at (-1.5,0.0) {};
\node[proc] (q1) at (1.5,0.0) {};
\draw (a0) -- (b1);
\node at (0,-0.75) {(After)};
\end{scope}
\end{tikzpicture}
\caption{An example of two-way linking.}
\label{fig:2-way-linking}
\end{figure}

\cref{fig:2-way-linking} shows an example of two-way linking. Rectangles
are threads, lines are channels, circles at both ends of a channel are
its endpoints, and the number in an endpoint is the role to be played by
the party holding the endpoint. In a binary session, there are only two
roles, abstracted as 0 and 1. The middle thread performs a link in
(Before), and the result is shown in (After) where the linking thread is
removed from the middle. Linking may look like an unfamiliar feature as
it is usually not seen in a message passing system that is not based on
session types. This feature enables the composition of sessions in a
well-defined way. In particular, one can only link two channels of the
\emph{same} session type, by connecting two \emph{dual} endpoints and
leaving the other two \emph{dual} endpoints communicating directly as if
they were the two endpoints of a newly formed channel.

While synchronous two-way linking may seem trivial to implement, any
practical implementation of two-way linking is inherently asynchronous,
where channels are buffered and sending on such channels is
non-blocking. Indeed asynchrony makes it difficult to merge two channels
as there are potentially unreceived messages left in the buffers
attached to them. Concurrent C0 \cite{Willsey:2016kz} is a notable
asynchronous implementation for a binary session type system that
supports two-way linking. In a \emph{binary} session, a channel is
shared by exactly two parties/participants. This fact can be used to
infer the direction of messages, which in turn indicates that only one
channel may have unreceived messages. However, there is no such
inference in the setting of multiparty sessions.

Multiparty session types \cite{Honda:2008hi} are introduced to specify
sessions involving more than two participants. While the original work
of \cite{Honda:2008hi} connects all the parties via a vector of
point-to-point channels, we instead use a single channel to connect all
endpoints here, essentially making the channel a message-bus or
\emph{blackboard}. In a multiparty session, all parties other than the
ones involved in a link can be simultaneously writing/reading on the
channel. Linking becomes completely symmetric, and there is not a single
unique direction of message flows. Also, the two channels being linked
may both contain unreceived messages, making it difficult to merge them
while preserving message orders. Based on our formulation of multiparty
session types \cite{Wu:2018tt}, a well-defined two-way linking with
\emph{residual}, or even \emph{three-way} linking, is also possible.
These features are essential in composing multiparty sessions
\cite{Xi:2016tr,Scalas:2017ci} but they are even more difficult to be
implemented correctly.

\begin{figure}
\begin{tikzpicture}
\tikzstyle{ep}=[circle,draw,minimum size=0.5cm,inner sep=0]
\tikzstyle{proc}=[rectangle,draw,minimum size=0.6cm,inner sep=0]
\begin{scope}
\node[ep] (a0) at (-1.5, 0.0) {0};
\node[ep] (a1) at (-0.5, 0.0) {1,2};
\node[ep] (b0) at (0.5, 0.0) {0,1};
\node[ep] (b1) at (1.5, 0.0) {2};
\draw (a0) -- (a1);
\draw (b0) -- (b1);
\node[proc] (p0) at (-1.5,0.0) {};
\node[proc] (q1) at (1.5,0.0) {};
\draw (-0.8,0.3) rectangle (0.8,-0.3);
\node [above] at (0,0.25) {$\dcut$};
\node at (0,-0.75) {(Before)};
\end{scope}

\begin{scope}[xshift=4cm]
\node[ep] (a0) at (-1.5, 0.0) {0};
\node[ep] (a1) at (0.0, 0.0) {1};
\node[ep] (a2) at (1.5, 0.0) {2};
\node[proc] (p0) at (-1.5,0.0) {};
\node[proc] (p1) at (0,0) {};
\node[proc] (p2) at (1.5,0.0) {};
\draw ([yshift=0.5cm]a0.north) -- ([yshift=0.5cm]a1.north) -- ([yshift=0.5cm]a2.north);
\draw (a0.north) -- ([yshift=0.5cm]a0.north);
\draw (a1.north) -- ([yshift=0.5cm]a1.north);
\draw (a2.north) -- ([yshift=0.5cm]a2.north);
\node at (0,-0.75) {(After)};
\end{scope}
\end{tikzpicture}
\caption{An example of two-way linking with residual.}
\label{fig:2-way-residual}
\end{figure}
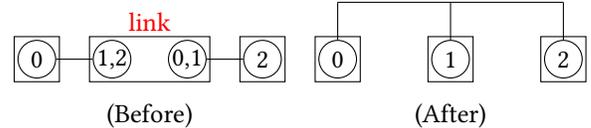

We present an example of a three-player (denoted by the roles 0, 1, and
2) game similar to the one in \cite{Scalas:2017ci} in
\cref{fig:2-way-residual}. Suppose player 1 (in the middle) would like
to initiate the game but does not know the other players yet. When
player 0 (on the left) comes, player 1 creates a channel, passing the
endpoint of role 0 to player 0, while player 1 holds the dual/complement
endpoint of roles 1 \emph{and} 2. Similarly, player 1 gives endpoint 2
to player 2 while holding complement endpoint of roles 0 \emph{and} 1.
Now, to start the game, player 1 can perform a two-way linking with
\emph{residual}\footnote{It relates to the logical rule
  \rulename{2-cut-residual} in multirole logic \cite{Xi:2017wv}.} by
merging the two endpoints that it holds into a \emph{single endpoint
with residual roles}. In this case, the residual role is the
intersection of \(\{0,1\}\cap\{1,2\}\), which equals \(\{1\}\). Please
see \cite{Xi:2017wv,Xi:2016tr,Wu:2018tt} for justification of the
correctness of this initialization process based on multirole logic.

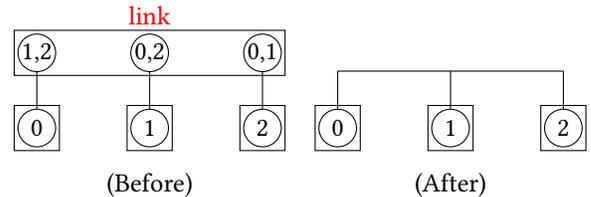
\begin{figure}
\begin{tikzpicture}
\tikzstyle{ep}=[circle,draw,minimum size=0.5cm,inner sep=0]
\tikzstyle{proc}=[rectangle,draw,minimum size=0.6cm,inner sep=0]
\begin{scope}
\node[ep] (a0) at (-1.5, -1.0) {0};
\node[ep] (a1) at (-1.5, 0.0) {1,2};
\node[ep] (b0) at (0.0, -1.0) {1};
\node[ep] (b1) at (0.0, 0.0) {0,2};
\node[ep] (c0) at (1.5,-1.0) {2};
\node[ep] (c1) at (1.5,0) {0,1};
\draw (a0) -- (a1);
\draw (b0) -- (b1);
\draw (c0) -- (c1);
\node[proc] (p0) at (a0) {};
\node[proc] (p1) at (b0) {};
\node[proc] (p2) at (c0) {};
\draw (-1.8,0.3) rectangle (1.8,-0.3);
\node [above] at (0,0.25) {$\dcut$};
\node at (0,-1.75) {(Before)};
\end{scope}

\begin{scope}[xshift=4cm, yshift=-1cm]
\node[ep] (a0) at (-1.5, 0.0) {0};
\node[ep] (a1) at (0.0, 0.0) {1};
\node[ep] (a2) at (1.5, 0.0) {2};
\node[proc] (p0) at (-1.5,0.0) {};
\node[proc] (p1) at (0,0) {};
\node[proc] (p2) at (1.5,0.0) {};
\draw ([yshift=0.5cm]a0.north) -- ([yshift=0.5cm]a1.north) -- ([yshift=0.5cm]a2.north);
\draw (a0.north) -- ([yshift=0.5cm]a0.north);
\draw (a1.north) -- ([yshift=0.5cm]a1.north);
\draw (a2.north) -- ([yshift=0.5cm]a2.north);
\node at (0,-0.75) {(After)};
\end{scope}
\end{tikzpicture}
\caption{An example of three-way linking.}
\label{fig:3-way-linking}
\end{figure}

Alternatively, we can also rely on a dedicated game server to match a
game for players (that do not know each other). Instead of relying on a
particular player, we can use a \emph{three-way} linking as in
\cref{fig:3-way-linking}. One can also perform two consecutive two-way
linking with residuals to achieve the same goal as a single three-way
linking.\footnote{The correctness of doing so is justified in multirole
  logic.} We focus on the implementation of two-way linking with
residual here as it is more general. Three-way linking can be
implemented either as two applications of two-way linking with residual,
or implemented directly following the same principle. Please find the
implementation code online
\url{https://github.com/steinwaywhw/ats-sessions}.

\hypertarget{runtime-implementation}{%
\section{Runtime Implementation}\label{runtime-implementation}}

Any multiparty session type system should come with, besides the type
system itself, a runtime that implements channels and operations on
channels, e.g. \texttt{send}, \texttt{receive}, and \texttt{link}. We
briefly describe such an asynchronous runtime.

Channels are implemented as a blackboard, where any party can read
messages that any other party writes. When implemented locally, the
blackboard can be a shared buffer. When implemented distributedly, the
blackboard can be a database. We abstract over this detail, only
assuming the following properties. First, the blackboard is unbounded in
capacity. Second, the blackboard should support atomic writes and atomic
\emph{selective} reads in the style of Erlang \cite{Armstrong:2003uc}.
Selective receive is essential for guaranteeing the order of the
received messages when there are multiple readers, writers, and kinds of
messages. Third, the blackboard preserves the order of messages.

A message consists of a header and a body, where the header contains a
label (denoting the kind of the message), the sender's role, and the
receivers' roles. For instance, we may use \texttt{MSG} for
synchronizing \texttt{send}/\texttt{receive}, \texttt{BRANCH} for
synchronizing \texttt{choose}/\texttt{offer}, etc. Also, we use
\texttt{KILL} and \texttt{KEEP} for linking. These header fields are
essential. When combined with selective reads, they can guarantee the
correct ordering of message exchanges. For instance, suppose both party
0 and party 2 send point-to-point messages of some label to party 1
asynchronously, then they may be written to the board in an unspecified
order. Therefore simply returning the first message is not correct. We
need to use selective receive based on the header in order to let party
1 deterministically retrieve the message based on the session type.
Essentially, the combination of message headers and selective receive
provides each endpoint a \emph{filtered view} of the board, where only
messages relevant to a party is present, and they are correctly ordered.
Please note that message ordering is not an issue in a binary session.
One only needs to guarantee a party does not read a message from the
board that is just written by itself. For instance, Concurrent C0
\cite{Willsey:2016kz} uses a direction flag for this purpose.

The receivers field of a message is also used for recording which party
is yet to receive the message. For instance, a thread may hold an
endpoint of roles \texttt{\{0,1\}}. If the thread receives a
point-to-point message for party \texttt{1}, then only \texttt{1} will
be removed from the receivers field of the message to mark it as read.
If the thread receives a broadcast message, then both \texttt{0} and
\texttt{1} will be removed. After all receiving parties have received
the message, the receivers field becomes empty and thus can be removed
from the blackboard. The receivers field of \texttt{KEEP}/\texttt{KILL}
has particular usages and is not subject to ``mark-as-read.''

The board provides, amongst others, two low-level APIs, \texttt{read}
and \texttt{write}. Informally, \texttt{read} has a signature of
\texttt{(label,\ sender\ role,\ receiver\ roles)\ -\textgreater{}\ payload}
and \texttt{write} has
\texttt{(label,\ sender\ role,\ receiver\ roles,\ payload)\ -\textgreater{}\ void}.
In our formulation of session types, the sender is always a single role,
while receivers can be either a single role for point-to-point
messaging, or the full set of roles w.r.t a session for broadcasting.
When invoking \texttt{read}, one needs to specify the receiver(s) to
work with ``mark-as-read.'' For instance, if a thread uses an endpoint
to receive a broadcast message, it should invoke \texttt{read} with all
the roles played by the endpoint. Note that \texttt{read} is selective,
and we define a match as follows. Given a pattern, a message is a match
if 1) the label is the same as the pattern, 2) the sender is the same as
the pattern, 3) the receivers are a superset of that in the pattern. We
say ``match'' from now on if these conditions are true. For
\texttt{KEEP}/\texttt{KILL}, we check for a match only based on
receivers and ignore labels and senders.

Each endpoint is a tuple of
\texttt{(roles,\ roles,\ reference\ to\ the\ board)} plus a set of
high-level APIs like \texttt{send}, \texttt{receive}, and \texttt{link}
implemented using low-level ones like \texttt{read}/\texttt{write} of
the board. The first field records the full set of roles w.r.t a
session, while the second field is the subset of roles played by the
endpoint. The thread holding the endpoint essentially plays these roles
within the session. For instance, a thread holding endpoint
\texttt{(\{0,1,2\},\{0\},Board\ 1)} plays role 0 in a three-party
session, where \texttt{Board\ 1} is a pointer/reference to some
shared-memory blackboard.

Blackboard is reference counted. Notably, the
\texttt{KEEP}/\texttt{KILL} message also contains a counted reference to
a blackboard that we shall detail later. When the reference count
decreases to zero, the blackboard will be freed. Because of reference
counting, session termination can be implemented asynchronously by
allowing each endpoint to terminate on its own, as compared to a
synchronized termination using a pair of functions like
\texttt{close}/\texttt{wait}. Note again that a binary session does not
need reference counted channels since it is always known to have exactly
two parties in a session.

\hypertarget{two-way-linking-in-multiparty-sessions}{%
\section{Two-way Linking in Multiparty
Sessions}\label{two-way-linking-in-multiparty-sessions}}

\begin{figure}
\begin{tikzpicture}
\tikzstyle{ep}=[circle,draw,minimum size=0.5cm,inner sep=0]
\tikzstyle{proc}=[rectangle,draw,minimum size=0.6cm,inner sep=0]
\tikzstyle{msg}=[rectangle,inner sep=0.1cm,outer sep=0.2cm]
\tikzstyle{board}=[rectangle,draw,anchor=north,inner sep=0,minimum width=3.5cm,minimum height=1cm]
\begin{scope}[xshift=-2cm]
\node[board] (b1) at (0,0) {};
\node at ([yshift=-0.3cm]b1.south) {Board 1};
\node[msg] (m1) at ([yshift=-0.3cm]b1.north) {[MSG] [f:t] payload};
\node[msg] (m2) at (m1.south) {[MSG] [f:t] payload};
\node[ep] (e1) at (-1,0.5) {0};
\node[ep] (e2) at (1,0.5) {1,2};
\node[proc] (p1) at (e1) {};
\draw (e1) -- (b1.north);
\draw (e2) -- (b1.north);
\end{scope}

\draw (-1.3,0.8) rectangle (1.3,0.2);
\node at (0,1.1) {$\dcut$}; 

\begin{scope}[xshift=2cm]
\node[board] (b1) at (0,0) {};
\node at ([yshift=-0.3cm]b1.south) {Board 2};
\node[msg] (m1) at ([yshift=-0.3cm]b1.north) {[MSG] [f:t] payload};
\node[ep] (e3) at (-1,0.5) {0,1};
\node[ep] (e4) at (1,0.5) {2};
\node[proc] (p2) at (e4) {};
\draw (e3) -- (b1.north);
\draw (e4) -- (b1.north);
\end{scope}
\end{tikzpicture}
\caption{Before linking.}
\label{fig:linking-1}
\end{figure}
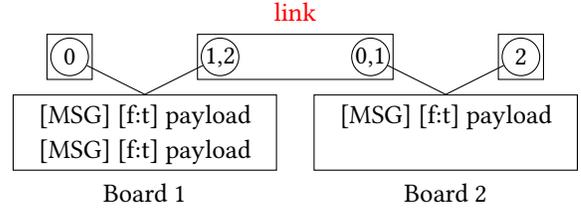

We describe the implementation of two-way linking in multiparty
sessions. We start with two blackboards being linked as in
\cref{fig:linking-1}, each containing some messages unreceived. For
instance, \texttt{{[}MSG{]}\ {[}f:t{]}\ payload} means the message is of
label \texttt{MSG}, with senders \texttt{f} (from), receivers \texttt{t}
(to), and some payload. An important difference from linking in binary
sessions is that both boards may contain messages needed by endpoints
from the other boards. For instance, there may be messages needed by
party 1 on both boards.

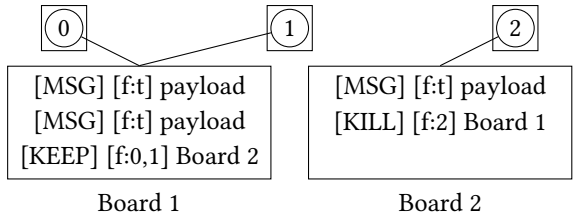
\begin{figure}
\begin{tikzpicture}
\tikzstyle{ep}=[circle,draw,minimum size=0.5cm,inner sep=0]
\tikzstyle{proc}=[rectangle,draw,minimum size=0.6cm,inner sep=0]
\tikzstyle{msg}=[rectangle,inner sep=0.1cm,outer sep=0.2cm]
\tikzstyle{board}=[rectangle,draw,anchor=north,inner sep=0,minimum width=3.5cm,minimum height=1.5cm]
\begin{scope}[xshift=-2cm]
\node[board] (b1) at (0,0) {};
\node at ([yshift=-0.3cm]b1.south) {Board 1};
\node[msg] (m1) at ([yshift=-0.3cm]b1.north) {[MSG] [f:t] payload};
\node[msg] (m2) at (m1.south) {[MSG] [f:t] payload};
\node[msg] (m3) at (m2.south) {[KEEP] [f:0,1] Board 2};
\node[ep] (e1) at (-1,0.5) {0};
\node[proc] (p1) at (e1) {};
\draw (e1) -- (b1.north);
\end{scope}
\node[ep] (e2) at (0,0.5) {1};
\node[proc] (p2) at (e2) {};
\draw (e2) -- (b1.north);

\begin{scope}[xshift=2cm]
\node[board] (b2) at (0,0) {};
\node at ([yshift=-0.3cm]b2.south) {Board 2};
\node[msg] (m1) at ([yshift=-0.3cm]b2.north) {[MSG] [f:t] payload};
\node[msg] (m2) at (m1.south) {[KILL] [f:2] Board 1};
\node[ep] (e4) at (1,0.5) {2};
\node[proc] (p2) at (e4) {};
\draw (e4) -- (b2.north);
\end{scope}
\end{tikzpicture}
\caption{Start linking.}
\label{fig:linking-2}
\end{figure}

To merge two boards into one, we drain one board until it has no
messages left, and reuse the other board as the resulting board. Since
linking is entirely symmetric, we randomly pick a board as the
\emph{keep} board, and the other as a \emph{kill} board. Let's assume
board 1 is the keep board, and board 2 is the kill board. We write a
message ``\texttt{{[}KEEP{]}\ {[}f:0,1{]}\ Board\ 2}'' to the keep
board, and a message ``\texttt{{[}KILL{]}\ {[}f:2{]}\ Board\ 1}'' to the
kill board where the receivers of both messages are essentially the
roles not involved in the link in their respective sessions, except that
\texttt{KEEP} additionally contains the residual roles. The receivers
field is especially important for avoiding \emph{self-loops} and these
roles are justified by session typing and LMRL. The sender fields are
not used. Both messages have \emph{counted} references to the other
board as their payloads. In the meantime, the middle thread will obtain
an endpoint (referencing the keep board) with residual roles. If the
residual roles are empty, the endpoint can be immediately closed.
\cref{fig:linking-2} shows what it looks like right after the linking
function returns.

\begin{figure}
\begin{tikzpicture}
\tikzstyle{ep}=[circle,draw,minimum size=0.5cm,inner sep=0]
\tikzstyle{proc}=[rectangle,draw,minimum size=0.6cm,inner sep=0]
\tikzstyle{msg}=[rectangle,inner sep=0.1cm,outer sep=0.2cm]
\tikzstyle{board}=[rectangle,draw,anchor=north,inner sep=0,minimum width=2.5cm,minimum height=1cm]
\begin{scope}
\node[board] (b1) at (0,0) {};
\node at ([yshift=-0.3cm]b1.south) {Board 1};
\node[msg] (m1) at ([yshift=-0.3cm]b1.north) {[KEEP] Board 2};
\node[msg] (m2) at (m1.south) {[KEEP] Board 3};
\end{scope}

\begin{scope}[xshift=2.75cm]
\node[board] (b2) at (0,0) {};
\node at ([yshift=-0.3cm]b2.south) {Board 2};
\node[msg] (m1) at ([yshift=-0.3cm]b2.north) {[KILL] Board 1};
\end{scope}

\begin{scope}[xshift=5.5cm]
\node[board] (b3) at (0,0) {};
\node at ([yshift=-0.3cm]b3.south) {Board 3};
\node[msg] (m1) at ([yshift=-0.3cm]b3.north) {[KILL] Board 1};
\end{scope}
\end{tikzpicture}
\caption{Chaining.}
\label{fig:chaining}
\end{figure}
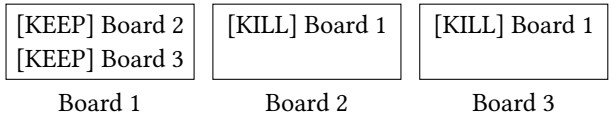

A crucial invariant is that \texttt{KILL} \emph{should be the last
message in a board}. Specifically, \texttt{write} follows \texttt{KILL},
but ignores \texttt{KEEP}. Namely, \texttt{write} appends to the end of
a destination board after being redirected by potentially many
\texttt{KILL} messages. By session typing, messages on both boards that
come before \texttt{KEEP}/\texttt{KILL} have disjoint senders. Thus they
can be merged safely without breaking topological orders. With the above
invariant, the corresponding implementation of \texttt{write} ensures
that messages after \texttt{KEEP} are already properly merged. As a
result, the implementation of \texttt{read} simply needs to respect this
order. Specifically, \texttt{read} attempts to match any messages before
\texttt{KEEP} or \texttt{KILL} first. Otherwise, if \texttt{read} sees a
\texttt{KEEP}, it is redirected to the referenced board, i.e., board 2.
If \texttt{read} fails again on board 2, it restarts searching from the
message right after the \texttt{KEEP} on board 1. Additionally, if
\texttt{KILL} is the only message left on board 2, the corresponding
\texttt{KEEP} in board 1 is deleted since board 2 is no longer relevant.
In the other case where \texttt{read} sees a \texttt{KILL}, it is
redirected to the referenced board, i.e., board 1, if \texttt{KILL} is a
match. Otherwise \texttt{read} fails if \texttt{KILL} is not a match,
which only happens when the \texttt{read} is redirected by a
corresponding \texttt{KEEP}. Reference counting ensures that boards are
safely freed eventually.

It is very common to have a long chain of linking, e.g., the queue
example given by \cite{Pfenning:2015hp}. It may result in a
configuration like \cref{fig:chaining}. The presented approach is
recursive and is valid in the presence of chaining. For instance,
\texttt{read} or \texttt{write} can be redirected multiple times.
Reference counting of boards guarantees that only when the board is
irrelevant to any endpoints that it can be safely freed. The decision to
put \emph{counted} references in \texttt{KEEP} and \texttt{KILL}
dramatically simplifies the implementation of linking in the presence of
chaining.

Interestingly, since three-way linking can be implemented using two
consecutive two-way linking with residual, chaining such as that in
\cref{fig:chaining} can be thought as a generalized three-way linking.
Namely, to implement three-way linking directly, one simply need to
insert two \texttt{KEEP} messages in the keep board, and one
\texttt{KILL} for each of the other two kill boards, just like
\cref{fig:chaining}.

\hypertarget{related-works-and-conclusions}{%
\section{Related Works and
Conclusions}\label{related-works-and-conclusions}}

The most related implementations of session type systems are SILL
\cite{Pfenning:2015hp} and Concurrent C0 \cite{Willsey:2016kz} based on
\cite{Caires:2010gi}, and Session Links \cite{Lindley:2015vy} based on
\cite{Wadler:2012ua}. Session Links does not support linking/forwarding.
SILL uses explicit forwarding to our best knowledge. Concurrent C0
implements linking by sending a \texttt{FWD}, which is also mentioned in
their recent work \cite{Pfenning:2018wk}. \texttt{FWD} is essentially
our \texttt{KILL}. Because a binary session only has two parties, it can
be shown based on session typing that the kill board will \emph{not}
have messages needed by parties referencing the keep board. Therefore
there is no need for a \texttt{KEEP} message to redirect \texttt{read}
to the kill board. With multiparty sessions, this inexplicit condition
no longer holds, and both boards need to reference each other. Our prior
work from late 2015 independently implemented linking in binary sessions
by writing a board reference to another board. Implementation wise, the
present paper draws inspirations from both our prior work and the work
from Concurrent C0. With the present work, the implementation of
Concurrent C0 can be thought as an optimized implementation for binary
sessions, where \texttt{KEEP} and reference counting are not needed, and
the linking thread always has empty residual roles allowing the thread
to be removed.

Another related work is \texttt{lchannels} in \cite{Scalas:2017ci}. In
their 3-player game example, a server creates a private 3-party session.
To start a game, the server sends out each endpoint to a player, via
private channels between the server and each player. This is formulated
as multiparty delegation/higher-order sessions. With our implementation,
this can be done directly by linking, avoiding those private channels,
and is arguably closer to real-world scenarios.
\cite{Caires:2016dn,Carbone:2016kd} uses arbiters, which are essentially
explicit forwarding.

To conclude, we generalized the implementation of linking to multiparty
sessions. We identified several additional requirements that are not
needed in binary sessions. One needs the \texttt{KEEP} in addition to
the \texttt{KILL}/\texttt{FWD} in order to redirect \texttt{read} from
the keep board to the kill board. One needs message headers and
selective receives for guaranteeing topological orders of messages. One
needs references counting to decide when to free channels. We also
identified two new primitives, two-way linking with residual and
three-way linking. To justify their correctness, we have to refer
readers to our prior work \cite{Xi:2017wv,Xi:2016tr,Wu:2018tt} due to
space limits. To our best knowledge, the two new kinds of linking are
novel, and the implementation of linking in multiparty session is also
novel.







\bibliography{library}


\begin{thebibliography}{17}


\ifx \showCODEN    \undefined \def \showCODEN     #1{\unskip}     \fi
\ifx \showDOI      \undefined \def \showDOI       #1{#1}\fi
\ifx \showISBNx    \undefined \def \showISBNx     #1{\unskip}     \fi
\ifx \showISBNxiii \undefined \def \showISBNxiii  #1{\unskip}     \fi
\ifx \showISSN     \undefined \def \showISSN      #1{\unskip}     \fi
\ifx \showLCCN     \undefined \def \showLCCN      #1{\unskip}     \fi
\ifx \shownote     \undefined \def \shownote      #1{#1}          \fi
\ifx \showarticletitle \undefined \def \showarticletitle #1{#1}   \fi
\ifx \showURL      \undefined \def \showURL       {\relax}        \fi
\providecommand\bibfield[2]{#2}
\providecommand\bibinfo[2]{#2}
\providecommand\natexlab[1]{#1}
\providecommand\showeprint[2][]{arXiv:#2}

\bibitem[\protect\citeauthoryear{Armstrong}{Armstrong}{2003}]%
        {Armstrong:2003uc}
\bibfield{author}{\bibinfo{person}{Joe Armstrong}.}
  \bibinfo{year}{2003}\natexlab{}.
\newblock {\em \bibinfo{title}{{Making reliable distributed systems in the
  presence of software errors.}}}
\newblock \bibinfo{thesistype}{Ph.D. Dissertation}.
\newblock
\showURL{%
\url{https://dblp.org/rec/phd/basesearch/Armstrong03}}


\bibitem[\protect\citeauthoryear{Caires and P{\'e}rez}{Caires and
  P{\'e}rez}{2016}]%
        {Caires:2016dn}
\bibfield{author}{\bibinfo{person}{Lu{\'\i}s Caires} {and}
  \bibinfo{person}{Jorge~A P{\'e}rez}.} \bibinfo{year}{2016}\natexlab{}.
\newblock \showarticletitle{{Multiparty Session Types Within a Canonical Binary
  Theory, and Beyond.}}
\newblock \bibinfo{journal}{{\em FORTE\/}} (\bibinfo{year}{2016}).
\newblock
\showDOI{%
\url{https://doi.org/10.1007/978-3-319-39570-8_6}}


\bibitem[\protect\citeauthoryear{Caires and Pfenning}{Caires and
  Pfenning}{2010}]%
        {Caires:2010gi}
\bibfield{author}{\bibinfo{person}{Lu{\'\i}s Caires} {and}
  \bibinfo{person}{Frank Pfenning}.} \bibinfo{year}{2010}\natexlab{}.
\newblock \showarticletitle{{Session Types as Intuitionistic Linear
  Propositions.}}. In \bibinfo{booktitle}{{\em CONCUR}}.
\newblock
\showDOI{%
\url{https://doi.org/10.1007/978-3-642-15375-4_16}}


\bibitem[\protect\citeauthoryear{Carbone, Lindley, Montesi, Sch{\"u}rmann, and
  Wadler}{Carbone et~al\mbox{.}}{2016}]%
        {Carbone:2016kd}
\bibfield{author}{\bibinfo{person}{Marco Carbone}, \bibinfo{person}{Sam
  Lindley}, \bibinfo{person}{Fabrizio Montesi}, \bibinfo{person}{Carsten
  Sch{\"u}rmann}, {and} \bibinfo{person}{Philip Wadler}.}
  \bibinfo{year}{2016}\natexlab{}.
\newblock \showarticletitle{{Coherence Generalises Duality - A Logical
  Explanation of Multiparty Session Types.}}
\newblock \bibinfo{journal}{{\em CONCUR\/}} (\bibinfo{year}{2016}).
\newblock
\showDOI{%
\url{https://doi.org/10.4230/LIPIcs.CONCUR.2016.33}}


\bibitem[\protect\citeauthoryear{Honda}{Honda}{1993}]%
        {Honda:1993eh}
\bibfield{author}{\bibinfo{person}{Kohei Honda}.}
  \bibinfo{year}{1993}\natexlab{}.
\newblock \showarticletitle{{Types for Dyadic Interaction.}}. In
  \bibinfo{booktitle}{{\em CONCUR}}.
\newblock
\showDOI{%
\url{https://doi.org/10.1007/3-540-57208-2_35}}


\bibitem[\protect\citeauthoryear{Honda, Vasconcelos, and Kubo}{Honda
  et~al\mbox{.}}{1998}]%
        {Honda:1998fm}
\bibfield{author}{\bibinfo{person}{Kohei Honda},
  \bibinfo{person}{Vasco~Thudichum Vasconcelos}, {and} \bibinfo{person}{Makoto
  Kubo}.} \bibinfo{year}{1998}\natexlab{}.
\newblock \showarticletitle{{Language Primitives and Type Discipline for
  Structured Communication-Based Programming.}}. In \bibinfo{booktitle}{{\em
  ESOP}}.
\newblock
\showDOI{%
\url{https://doi.org/10.1007/BFb0053567}}


\bibitem[\protect\citeauthoryear{Honda, Yoshida, and Carbone}{Honda
  et~al\mbox{.}}{2008}]%
        {Honda:2008hi}
\bibfield{author}{\bibinfo{person}{Kohei Honda}, \bibinfo{person}{Nobuko
  Yoshida}, {and} \bibinfo{person}{Marco Carbone}.}
  \bibinfo{year}{2008}\natexlab{}.
\newblock \showarticletitle{{Multiparty asynchronous session types.}}
\newblock \bibinfo{journal}{{\em POPL\/}} (\bibinfo{year}{2008}).
\newblock
\showDOI{%
\url{https://doi.org/10.1145/1328438.1328472}}


\bibitem[\protect\citeauthoryear{Lindley and Morris}{Lindley and
  Morris}{2015}]%
        {Lindley:2015vy}
\bibfield{author}{\bibinfo{person}{Sam Lindley} {and}
  \bibinfo{person}{J~Garrett Morris}.} \bibinfo{year}{2015}\natexlab{}.
\newblock \showarticletitle{{Lightweight Functional Session Types}}.
\newblock In \bibinfo{booktitle}{{\em Behavioural Types}}.
\newblock


\bibitem[\protect\citeauthoryear{Pfenning and Griffith}{Pfenning and
  Griffith}{2015}]%
        {Pfenning:2015hp}
\bibfield{author}{\bibinfo{person}{Frank Pfenning} {and}
  \bibinfo{person}{Dennis Griffith}.} \bibinfo{year}{2015}\natexlab{}.
\newblock \showarticletitle{{Polarized Substructural Session Types.}}. In
  \bibinfo{booktitle}{{\em FoSSaCS}}. \bibinfo{publisher}{Springer, Berlin,
  Heidelberg}, \bibinfo{pages}{3--22}.
\newblock
\showDOI{%
\url{https://doi.org/10.1007/978-3-662-46678-0_1}}


\bibitem[\protect\citeauthoryear{Pfenning and PRUIKSMA}{Pfenning and
  PRUIKSMA}{2018}]%
        {Pfenning:2018wk}
\bibfield{author}{\bibinfo{person}{F Pfenning} {and} \bibinfo{person}{K
  PRUIKSMA}.} \bibinfo{year}{2018}\natexlab{}.
\newblock \bibinfo{title}{{Asynchronous Multistructural Session Types}}.
\newblock   (\bibinfo{year}{2018}).
\newblock
\showURL{%
\url{http://www.cs.cmu.edu/~fp/papers/multi18.pdf}}


\bibitem[\protect\citeauthoryear{Scalas, Dardha, Hu, and Yoshida}{Scalas
  et~al\mbox{.}}{2017}]%
        {Scalas:2017ci}
\bibfield{author}{\bibinfo{person}{Alceste Scalas}, \bibinfo{person}{Ornela
  Dardha}, \bibinfo{person}{Raymond Hu}, {and} \bibinfo{person}{Nobuko
  Yoshida}.} \bibinfo{year}{2017}\natexlab{}.
\newblock \showarticletitle{{A Linear Decomposition of Multiparty Sessions for
  Safe Distributed Programming.}}. In \bibinfo{booktitle}{{\em ECOOP}}.
\newblock
\showDOI{%
\url{https://doi.org/10.4230/LIPIcs.ECOOP.2017.24}}


\bibitem[\protect\citeauthoryear{Takeuchi, Honda, and Kubo}{Takeuchi
  et~al\mbox{.}}{1994}]%
        {Takeuchi:1994bv}
\bibfield{author}{\bibinfo{person}{Kaku Takeuchi}, \bibinfo{person}{Kohei
  Honda}, {and} \bibinfo{person}{Makoto Kubo}.}
  \bibinfo{year}{1994}\natexlab{}.
\newblock \showarticletitle{{An Interaction-based Language and its Typing
  System.}}. In \bibinfo{booktitle}{{\em PARLE}}.
\newblock
\showDOI{%
\url{https://doi.org/10.1007/3-540-58184-7_118}}


\bibitem[\protect\citeauthoryear{Wadler}{Wadler}{2012}]%
        {Wadler:2012ua}
\bibfield{author}{\bibinfo{person}{Philip Wadler}.}
  \bibinfo{year}{2012}\natexlab{}.
\newblock \showarticletitle{{Propositions as sessions.}}
\newblock \bibinfo{journal}{{\em ICFP\/}} (\bibinfo{year}{2012}).
\newblock
\showDOI{%
\url{https://doi.org/10.1145/2364527.2364568}}


\bibitem[\protect\citeauthoryear{Willsey, Prabhu, and Pfenning}{Willsey
  et~al\mbox{.}}{2016}]%
        {Willsey:2016kz}
\bibfield{author}{\bibinfo{person}{Max Willsey}, \bibinfo{person}{Rokhini
  Prabhu}, {and} \bibinfo{person}{Frank Pfenning}.}
  \bibinfo{year}{2016}\natexlab{}.
\newblock \showarticletitle{{Design and Implementation of Concurrent C0.}}. In
  \bibinfo{booktitle}{{\em LINEARITY}}.
\newblock
\showDOI{%
\url{https://doi.org/10.4204/EPTCS.238.8}}


\bibitem[\protect\citeauthoryear{Wu and Xi}{Wu and Xi}{2018}]%
        {Wu:2018tt}
\bibfield{author}{\bibinfo{person}{Hanwen Wu} {and} \bibinfo{person}{Hongwei
  Xi}.} \bibinfo{year}{2018}\natexlab{}.
\newblock \showarticletitle{{Multiparty Dependent Session Types (Extended
  Abstract).}}
\newblock \bibinfo{journal}{{\em CoRR\/}} (\bibinfo{year}{2018}).
\newblock
\showURL{%
\url{https://dblp.org/rec/journals/corr/abs-1808-00077}}


\bibitem[\protect\citeauthoryear{Xi and Wu}{Xi and Wu}{2016}]%
        {Xi:2016tr}
\bibfield{author}{\bibinfo{person}{Hongwei Xi} {and} \bibinfo{person}{Hanwen
  Wu}.} \bibinfo{year}{2016}\natexlab{}.
\newblock \showarticletitle{{Linearly Typed Dyadic Group Sessions for Building
  Multiparty Sessions.}}
\newblock \bibinfo{journal}{{\em CoRR\/}} (\bibinfo{year}{2016}).
\newblock
\showURL{%
\url{http://dblp.org/rec/journals/corr/XiW16}}


\bibitem[\protect\citeauthoryear{Xi and Wu}{Xi and Wu}{2017}]%
        {Xi:2017wv}
\bibfield{author}{\bibinfo{person}{Hongwei Xi} {and} \bibinfo{person}{Hanwen
  Wu}.} \bibinfo{year}{2017}\natexlab{}.
\newblock \showarticletitle{{Multirole Logic (Extended Abstract).}}
\newblock \bibinfo{journal}{{\em CoRR\/}} (\bibinfo{year}{2017}).
\newblock
\showURL{%
\url{http://dblp.org/rec/journals/corr/XiW17}}


\end{thebibliography}

%
\end{document}